\documentclass[12pt]{article}
\usepackage{amssymb,amsmath,cite}
\numberwithin{equation}{section}
\begin{document}
\begin{titlepage}
      \rightline{CERN-TH/2003-279}
      \rightline{hep-th/0312230}
      \vskip 1cm
      \centerline{{\Large \bf A Note on UV/IR Mixing and}}
      \centerline{{\Large \bf Non-Com\-mu\-ta\-tive Instanton Calculus}}
      \vskip 1cm
      \centerline{Andreas A.~Bichl}
      \vskip 0.5cm
      \centerline{Theory Division, CERN}
      \centerline{1211 Geneva 23, Switzerland}
      \vskip 0.3cm
      \centerline{Andreas.Bichl@cern.ch}
      \vskip 1.5cm
      \begin{abstract}
            \noindent
            We estimate the instanton-induced vacuum energy in non-com\-mu\-ta\-tive $U(1)$ Yang--Mills
            theory in four dimensions. In the dilute gas approximation, it is found to be plagued by
            infrared divergences, as a result of UV/IR mixing.
      \end{abstract}
\end{titlepage}
\section{Introduction}
Non-com\-mu\-ta\-tive gauge theories have attracted a lot of attention in recent years, since they appeared as low-energy effective descriptions of open strings on a D-brane with a constant background $B$-field \cite{Seiberg:1999vs}. For general reviews on non-com\-mu\-ta\-tive field theories, see for instance \cite{Douglas:2001ba,Szabo:2001kg}. Perturbative studies of these theories revealed a very interesting feature---the so-called {\em UV/IR mixing\/} \cite{Minwalla:1999px,Matusis:2000jf,VanRaamsdonk:2000rr}.

All Feynman graphs in a NC field theory can be divided into {\em planar\/} and {\em non-planar\/} graphs \cite{Filk:dm}. The planar ones are equal to their commutative counterpart multiplied by a phase factor, which depends only on external momenta. From this sector we get the usual UV divergences, which are handled with standard renormalization techniques.

Non-planar graphs include phase factors like $\exp\,(\text{i}\:k\cdot\theta\cdot p)$, with $k$ the loop-momentum, $p$ an external momentum and $\theta$ the NC parameter. For very high loop momenta, the phase factor oscillates very fast and renders the integral finite. There are no UV divergences coming from the non-planar sector. However, this is only valid for a non-vanishing external momentum $\theta\cdot p$. Taking the limit $\theta\cdot p\rightarrow 0$ brings back the infinity, but this time as an IR divergence. Therefore, UV modes {\em do not decouple\/} from IR modes in non-com\-mu\-ta\-tive field theories.  

The aim of this paper is to study the impact of UV/IR mixing on the simplest non-perturbative effects, namely the contribution of a {\em dilute instanton gas\/} to the vacuum energy. Each non-com\-mu\-ta\-tive instanton or anti-instanton contributes to the vacuum-to-vacuum amplitude a factor proportional to $\exp\,(-8\pi^2/g^2\pm\text{i}\,\vartheta)$, where $g$ is the Yang--Mills coupling and $\vartheta$ is the topological angle.\footnote{We use the symbol $\theta$ for the NC parameter, and $\vartheta$ for the topological angle.} In the dilute gas approximation, standard treatment \cite{Coleman} yields
\begin{align}
   \Delta E_{\text{inst}} = -2 \; K/T \; \cos\vartheta\,,
\end{align}
for the $\vartheta$-dependent part of the vacuum energy, where $T$ is the Euclidean time extent of the four-dimensional box. The factor $K$ results from the contribution of collective coordinates, Jacobians, and perturbative corrections to the instanton background \cite{Coleman,'tHooft:fv,Vainshtein:wh}:
\begin{align}
   K \propto V \, \Lambda_{\text{UV}}^4 \, \left( \sqrt{S_{\text{cl}}} \right)^4 \,
   e^{-S_{\text{cl}}} \; e^{-S_{\text{eff}}({\cal{A}})} \, ,
   \label{zm}
\end{align}
where $V$ is the four-dimensional volume corresponding to the four position moduli of the instanton (the only modular parameters of the minimum charge instanton), $\Lambda_{\text{UV}}$ is an ultraviolet cutoff, $S_{\text{cl}}=8\pi^2/g^2$, and $S_{\text{eff}}$ is the perturbative effective action evaluated on the instanton field ${\cal{A}}$. To leading order, $S_{\text{eff}}$ is given by a ratio of determinants of the gauge field and ghost quadratic fluctuation operators.

In this note we study the infrared behaviour of the perturbative effective action $S_{\text{eff}}$ in the one-loop approximation. In normal gauge theories, the size of the instanton $\rho$ acts as an infrared cutoff, although subsequent integration over $\rho$ yields an infrared divergence over moduli space. In the present non-com\-mu\-ta\-tive case, instantons have a fixed size: $\rho\sim\sqrt{\theta}$. Despite this fact, we shall find that UV/IR mixing effects render the effective action infrared-divergent.

The paper is organized as follows: In section \ref{ncu1inst} we briefly review non-com\-mu\-ta\-tive gauge theories and the construction of non-com\-mu\-ta\-tive instantons. We will restrict ourselves to the simplest case of a NC $U(1)$ anti-selfdual instanton, which is sufficient to show all the basic features. Section \ref{1linstdet} is devoted to an estimate of the one-loop instanton determinant in NC $U(1)$ Yang--Mills theory, and we comment on the role of supersymmetry in this context. The last section presents a discussion of our results.
\section{The Non-Com\-mu\-ta\-tive $U(1)$ Instanton}\label{ncu1inst}
We consider a four-dimensional non-com\-mu\-ta\-tive Euclidean space, which is represented by coordinates $x_\mu$ obeying the following algebra:
\begin{align}
   \left[ x_{\mu},x_{\nu} \right] = \text{i} \theta_{\mu\nu},
\end{align}
where $\theta_{\mu\nu}$ is a constant antisymmetric matrix and $\mu,\nu=1,2,3,4$. This non-com\-mu\-ta\-tive structure of space-time is implemented via the Moyal product, given by
\begin{align}
   (f \star g) (x) = f(x) \: e^{\frac{\text{i}}{2}\theta_{\mu\nu} \:
   \overleftarrow{\partial_\mu} \overrightarrow{\partial_\nu}} g(x).
\end{align}
Using this $\star$-product, the Lagrangian of NC $U(1)$ Yang--Mills theory reads
\begin{align}
   S = -\frac{1}{2g^2} \int d^4x\; F_{\mu\nu} \star F_{\mu\nu}
   -\frac{\text{i}\,\vartheta}{8\pi^2} \int F \wedge F ,
   \label{S}
\end{align}
with the field strength
\begin{align}
   F_{\mu\nu} = \partial_\mu A_\nu - \partial_\nu A_\mu - \text{i}
   \left[ A_\mu , A_\nu \right]_\star.
\end{align}

Now, we proceed to briefly review the ADHM construction of instanton solutions in non-com\-mu\-ta\-tive gauge theories. This procedure describes an algebraic way of finding (anti-)selfdual configurations of the gauge field, which correspond to a certain class of classical solutions of the field equations, {\em i.e.~}instantons. The ADHM method was introduced by \cite{Atiyah:ri} and put forward to the non-com\-mu\-ta\-tive regime by \cite{Nekrasov:1998ss}. The existence of non-trivial solutions were shown even in the case of the gauge group $U(1)$, where commutative solutions are lacking. Since then, a lot of effort has been made in order to construct instantons and study their properties in various NC gauge theories (see for instance \cite{Nekrasov:2000ih} and references therein). Owing to Corrigan's identity it was shown that they have integer topological charge, like their commutative counterparts.

We will restrict ourselves to the Euclidean space $\mathbb{R}_{\text{NC}}^2\times\mathbb{R}^2$, which is given by the direct product of two-dimensional non-com\-mu\-ta\-tive space with two-dimensional ordinary space. This case corresponds to space--space non-commutativity\footnote{Using Euclidean space-time rotations, we can always transform a three-dimensional non-com\-mu\-ta\-tive space into the product space $\mathbb{R}_{\text{NC}}^2\times\mathbb{R}^2$.} and therefore avoids unitarity problems of the associated Lorentzian theory \cite{Gomis:2000zz,Alvarez-Gaume:2001ka}.

Parametrizing the space $\mathbb{R}_{\text{NC}}^2\times\mathbb{R}^2$ via $(x_1,x_2) \times (x_3,x_4)$, where
\begin{align}
   \theta_{12}=-\theta_{21}=\theta=\zeta /2,
   \label{t}
\end{align}
and passing to complex coordinates,
\begin{alignat}{2}
   z_1 &= x_2+\text{i}x_1, \qquad & \bar{z}_1 &= x_2-\text{i}x_1, \nonumber \\
   z_2 &= x_4+\text{i}x_3, \qquad & \bar{z}_2 &= x_4-\text{i}x_3,
   \label{cc}
\end{alignat}
we end up with the following commutator relations:
\begin{align}
   \left[ \bar{z}_1,z_1 \right] = \zeta,
   \quad
   \left[ \bar{z}_2,z_2 \right] = 0,
   \quad
   \left[ z_i,z_j \right] = 0.
\end{align}
We easily realize the coordinates $\bar{z}_1$ and $z_1$ as an annihilation and a creation operator acting in a Fock space ${\cal{H}}$ for a simple harmonic oscillator spanned by a basis $\left| n \right>$ with $n\ge 0$:
\begin{align}
   \bar{z}_1 \left| n \right> = \sqrt{\zeta n} \left| n-1 \right>, \qquad
   z_1 \left| n \right> = \sqrt{\zeta (n+1)} \left| n+1 \right>.
\end{align}
On the other hand, the coordinates $\bar{z}_2$ and $z_2$ are still ordinary c-numbers. Therefore, all fields on the space $\mathbb{R}_{\text{NC}}^2\times\mathbb{R}^2$ will be described by operator-valued expressions on the non-com\-mu\-ta\-tive plane ($\bar{z}_1$,$z_1$) and ordinary functions on the commutative plane ($\bar{z}_2$,$z_2$).

Using this preparation of the configuration space we can solve the deformed ADHM equations, which will lead us to the wanted instanton solution. For further details, we refer the reader to the literature \cite{Furuuchi:1999kv,Kim:2000ms,Furuuchi:2000vc,Chu:2001cx,Kim:2001ai,Kraus:2001xt}, where this topic has been extensively discussed.

Here, we will start from a solution for the anti-selfdual $U(1)$ instanton obtained in \cite{Kim:2001ai}:
\begin{align}
   A = \psi^\dagger d \psi,
\end{align}
where they introduce the exterior derivative $d=dz_i\partial_i+d\bar{z}_i\bar{\partial}_i\;(i=1,2)$, with $\partial_i=\partial/\partial z_i$, and use the language of differential forms.\footnote{Here, $dz_i$ and $d\bar{z}_i$ are anticommuting differentials. They commute with $z_i$ and $\bar{z}_i$ \cite{Nekrasov:1998ss}.} The vector $\psi$ reads\footnote{The vector $\psi$ is the zero mode of the Dirac operator $D^{\dagger}$, which is given by a certain matrix in the ADHM construction \cite{Kim:2001ai}.}
\begin{gather}
   \psi = \left( \begin{array}{c} \psi_1 \\ \psi_2 \\ \xi \end{array} \right)
   \quad \text{with} \quad
   \psi_1 = \bar{z}_2 \sqrt{\frac{\zeta}{\delta\Delta}},
   \quad \psi_2 = \bar{z}_1 \sqrt{\frac{\zeta}{\delta\Delta}},
   \quad \xi = \sqrt{\frac{\delta}{\Delta}},
   \nonumber
\\
   \quad \delta=z_1\bar{z}_1+z_2\bar{z}_2,
   \quad \Delta=\delta+\zeta,
   \quad \nabla=\delta-\zeta.   
\end{gather}
After a straightforward but tedious calculation, where we have to keep control over the operator ordering in the $(\bar{z}_1,z_1)$-plane:
\begin{xalignat}{2}
   \bar{z}_1 f(\delta) &= f(\Delta) \: \bar{z}_1, &
   z_1 f(\delta) &= f(\nabla) \: \bar{z}_1,
   \nonumber
\\
   \partial_1f &= \zeta^{-1} \left[ \bar{z}_1,f \right], &
   \bar{\partial}_1f &= -\zeta^{-1} \left[ z_1,f \right],
   \nonumber
\\
   \partial_1f^{-1} &= -f^{-1} \left( \partial_1f \right) f^{-1}, &
   \bar{\partial}_1f^{-1} &= -f^{-1} \left( \bar{\partial}_1f \right) f^{-1},
\end{xalignat}
we are able to derive an explicit expression for the instanton:
\begin{align}
   A &= \frac{1}{\zeta} \left[ \frac{\sqrt{(\Delta+\zeta)\delta}}{\Delta}-1 \right] \bar{z}_1 dz_1
   + \frac{1}{\zeta} \left[ 1-\frac{\sqrt{\Delta\nabla}}{\delta} \right] z_1 d\bar{z}_1 \nonumber
\\
   &- \left[ \frac{\zeta}{2\Delta\delta} \right] \left( \bar{z}_2 dz_2 - z_2 d\bar{z}_2 \right).
   \label{A}
\end{align}

Before converting this expression further, let us check if it really corresponds to an anti-selfdual solution with topological charge $-1$. Using all the above ingredients we obtain, for the corresponding field strength $F=dA+A\wedge A$:
\begin{align}
   F &= \frac{\zeta}{\Delta^2\delta^2} \left( \Delta z_1\bar{z}_1-\delta z_2\bar{z}_2 \right)
   \left( dz_2 d\bar{z}_2 - dz_1 d\bar{z}_1 \right) \nonumber
\\
   &+ \frac{2\zeta}{\Delta^2\sqrt{(\Delta+\zeta)\delta}} z_2 \bar{z}_1 d\bar{z}_2 dz_1
   + \frac{2\zeta}{\delta^2\sqrt{\Delta\nabla}} z_1 \bar{z}_2 d\bar{z}_1 dz_2,
\end{align}
which is indeed anti-selfdual, because $\ast F=-F$ implies $F_{z_1\bar{z}_1}+F_{z_2\bar{z}_2}=0$ and $F_{z_1z_2}=F_{\bar{z}_1\bar{z}_2}=0$.\footnote{The reader should not confuse the Hodge operator $\ast$ with the Moyal product $\star$.} Calculating the topological charge
\begin{align}
   Q = -\frac{1}{8\pi^2} \int F \wedge F,
\end{align}
via mapping of the integration over the $(\bar{z}_1,z_1)$-plane to the trace in the Fock space ${\cal{H}}$ (see \cite{Kim:2001ai})
\begin{align}
   \int d^4z\; {\cal{O}} (z) \;\longrightarrow\; -8\pi^2\zeta \sum_{n\ge 0}
   \int_0^\infty \kappa\;d\kappa\; \left< n \right| {\cal{O}} (z) \left| n \right>,
\end{align}
with $\kappa^2=x_3^2+x_4^2$, leads to the expected result $Q=-1$.

Established the expression (\ref{A}) as the anti-selfdual one instanton solution of NC $U(1)$ YM theory in the space $\mathbb{R}_{\text{NC}}^2\times\mathbb{R}^2$, we can proceed in taking a closer look at the infrared tail of the instanton, which will play the important role of our considerations in the next section. This can be achived very easily in recognizing the operator
\begin{align}
   \delta=z_1\bar{z}_1+z_2\bar{z}_2=x_1^2+x_2^2+x_3^2+x_4^2 \equiv r^2,
\end{align}
as the square of the distance $r$ from the origin in $\mathbb{R}_{\text{NC}}^2\times\mathbb{R}^2$. Taking the limit $\delta\rightarrow\infty$ of the instanton solution (\ref{A}) we get
\begin{align}
   \lim_{\delta\rightarrow\infty} A
   = \frac{\zeta}{2\delta^2}
   \left( -\bar{z}_1 dz_1 + z_1 d\bar{z}_1 - \bar{z}_2 dz_2 + z_2 d\bar{z}_2 \right).
   \label{Az}
\end{align}
Using the relations (\ref{t}) and (\ref{cc}) we can write (\ref{Az}) in the more convenient form
\begin{align}
   \lim_{r\rightarrow\infty} A_{\mu}(x)=-2\text{i}\,\theta\,\epsilon_{\mu\nu}\,\frac{x_{\nu}}{x^4},
   \label{instirx}
\end{align}
or by performing a Fourier transform
\begin{align}
   \lim_{p\rightarrow 0} A_{\mu}(p)=4\pi^2\,\theta\,\epsilon_{\mu\nu}\,\frac{p_{\nu}}{p^2},
   \label{instir}
\end{align}
where $\epsilon_{12}=\epsilon_{34}=1$ and $\epsilon_{21}=\epsilon_{43}=-1$.

Therefore, the non-com\-mu\-ta\-tive instanton (\ref{instirx}) decreases with $1/r^3$ very far from the origin and it can be shown that it has only significant values in a region $r\lesssim\sqrt{\theta}$. Furthermore, the instanton tail vanishes completely in the commutative limit $\theta\rightarrow 0$, but it will leave a singularity at the origin, which can be seen from (\ref{A}). This is consistent with our statement at the beginning of this section that there are no smooth instanton solutions in usual Maxwell theory (see \cite{Seiberg:1999vs,Kraus:2001xt} for further discussions on this point).
\section{The One-Loop Instanton Determinant}\label{1linstdet}
At the one-loop order, the effective action $S_{\text{eff}}({\cal{A}})$ is given by a ratio of determinants,
\begin{align}
   S_{\text{eff}} ({\cal{A}}) =
   - \frac{1}{2} \log \, {\det}' \, L_{\text{gauge}}
   + \log \, {\det}' \, L_{\text{ghost}} \,,
   \label{L}
\end{align}
of the quadratic fluctuation operators in the gauge and ghost sectors
\begin{align}
   (L_{\text{gauge}})_{\mu\nu} &= \left( D_\rho \star D_\rho \right) \delta_{\mu\nu}
   -2\text{i} \, F_{\mu\nu} - \left( 1 - \frac{1}{\alpha} \right) D_\mu \star D_\nu,
\\
   L_{\text{ghost}} &= D_\rho \star D_\rho,
\end{align} 
where $D_\mu=\partial_\mu-\text{i}\left[A_\mu,\;\;\right]_\star$ and $\alpha$ denotes the gauge parameter. Furthermore, ${\det}'$ indicates that the zero modes have to be omitted when computing the determinants. The complete effective action, including the $\log \, {\det}'$ contributions and the zero mode contributions can be written as a formal sum over all one-loop diagrams with external legs on the classical instanton profile ${\cal{A}}$. In this representation, the zero mode terms should arise as convenient resummations of infrared divergences to all orders.

At the level of planar diagrams we have a situation entirely similar to that of ordinary $SU(N)$ gauge theory, in the formal limit $N\rightarrow 1$. For example, the logarithmic dependence on the ultraviolet cutoff $\Lambda_{\text{UV}}$ arises from the planar two-point function \cite{Vainshtein:wh} and combines with the explicit dependence from the zero modes (\ref{zm}) to produce the usual renormalization group invariant expression
\begin{align}
   \Lambda_{\text{UV}}^4 \, \exp \left[ - \frac{8\pi^2}{g^2} + (\beta_0-4)
   \, \log \, (\rho\,\Lambda_{\text{UV}}) +
   \ldots \right] \;\longrightarrow\; \rho^{-4} \, (\rho\,\Lambda)^{\beta_0},
   \label{rzm}
\end{align} 
where $\beta_0$ is the one-loop beta-function coefficient (equal to $11/3$ for the case of pure NC $U(1)$ YM). The size of the instanton $\rho\sim\sqrt{\theta}$ acts like an infrared cutoff, since the classical field ${\cal{A}}$ decays to zero on distances larger than the instanton size. The dynamical scale $\Lambda$ is given by
\begin{align}
   \Lambda = \Lambda_{\text{UV}} \, \exp \left( -8\pi^2/\beta_0g^2 \right).
\end{align}
Hence, even if we cannot calculate the numerical coefficients in a precise way, a combination of dimensional analysis and the general properties of the perturbative effective action allows us to determine the gross features of the planar contribution to the instanton measure.

The dots in Eq.~(\ref{rzm}) stand for other UV-finite perturbative contributions. Among those, the non-planar diagrams of low order have strong IR singularities as a result of the famous UV/IR mixing effects. Despite the fact that the instanton profile vanishes at long distances, we must then check the infrared behaviour of the one-loop effective action. Here, in the non-planar sector, we do not have to implement the zero modes (corresponding to a summation over an infinite number of diagrams), because we focus on peculiar IR singularities that arise only from a finite number of diagrams, which is enough to estimate their effect. We will split the analysis into two parts: we will consider the IR poles (I) and the IR logarithms (II) separately. 
\subsection{The IR Pole Structure}
The pole structure of non-planar $n$-point functions can be read off the following gauge-invariant expression for the effective action \cite{Armoni:2001uw} of pure NC $U(1)$ YM
\begin{align}
   S^{\text{I}}_{\text{eff}} ({\cal{A}}) =
   \frac{1}{2\pi^2}
   \int \frac{d^4p}{(2\pi)^4} \; W'(-p) \;\frac{p^2}{\tilde p^2} \; K_2(\sqrt{p^2\tilde{p}^2})\; W'(p),
\label{p}
\end{align}
where $W'(p)$ denotes a truncated open Wilson line operator. It is worth noticing here that we have to make use of such operators in order to write down gauge-invariant expressions in non-com\-mu\-ta\-tive field theories.

Because we are mainly interested in the IR regime of the theory we will expand the modified Bessel function $K_2(z)$ for small momenta
\begin{align}
   \frac{p^2}{\tilde p^2} \; K_2(\sqrt{p^2\tilde{p}^2}) =
   \frac{2}{\tilde p^4} - \frac{p^2}{2\tilde p^2} + {\cal{O}}(\tilde p^0).
\label{K}
\end{align}
Insertion of the Wilson line operators (given in \cite{Armoni:2001uw}) in the IR regime, with ${\cal{A}}$ denoting the classical background gauge field
\begin{align}
   W'(p) =
   \text{i} \; \tilde p^\mu {\cal{A}}_\mu (p) - 
   \frac{1}{2} \int \frac{d^4 q}{(2 \pi)^4} \;
   \tilde p^\mu \tilde p^\nu 
   {\cal{A}}_\mu (p-q) {\cal{A}}_\nu(q) + \ldots,
\label{exp}
\end{align}
and performing a Wick rotation leads to the following Euclidean expressions for the two- and three-point functions
\begin{align}
   S^{\text{I},(2)}_{\text{eff}} ({\cal{A}}) &=
   \int \frac{d^4p}{(2\pi)^4} \; {\cal{A}}_{\mu}(p) {\cal{A}}_{\nu}(-p) \Pi_{\mu\nu}^{\text{I},(2)}(p),
   \label{2}
\\
   S^{\text{I},(3)}_{\text{eff}} ({\cal{A}})
   &= \int \frac{d^4p}{(2\pi)^4}\frac{d^4q}{(2\pi)^4} \;
   {\cal{A}}_{\mu}(p) {\cal{A}}_{\nu}(q) {\cal{A}}_{\rho}(-p-q) \Pi_{\mu\nu\rho}^{\text{I},(3)}(p),
   \label{3}
\end{align}
with
\begin{align}
   \Pi_{\mu\nu}^{\text{I},(2)}(p) &= -
   \frac{1}{\pi^2} \frac{\tilde{p}_{\mu}\tilde{p}_{\nu}}{\tilde{p}^4}
   +{\cal{O}}(\tilde{p}^0),
   \label{pii2}
\\
   \Pi_{\mu\nu\rho}^{\text{I},(3)}(p) &= - \frac{1}{\pi^2}
   \frac{\tilde{p}_\mu\tilde{p}_\nu\tilde{p}_\rho}{\tilde{p}^4}
   + {\cal{O}}(\tilde{p}^0).
\end{align}

It will be sufficient to consider these functions, since higher ones cannot lead to IR-divergent terms. Looking at the expansions (\ref{K}) and (\ref{exp}) we recognize that the $n$-point functions $\Pi^{\text{I},(n)}(p)$ will lead to pole structures of the order of $\tilde{p}^{(n-4)}$. Furthermore, we have to check the IR structure of our background gauge field ${\cal{A}}$, which will be used to calculate the contributions to the effective action. The instanton field (\ref{instir}) obtained previously will play this role. Every term of $S^{\text{I}}_{\text{eff}}$ has the following form (we skip Lorentz indices):
\begin{align}
   S^{\text{I},(n)}_{\text{eff}} \propto \!
   \int d^4p \, d^4q_1 \cdots d^4q_{n-2} \,
   {\cal{A}}(p) {\cal{A}}(q_1) \cdots {\cal{A}}(q_{n-2}) {\cal{A}}(-p-\textstyle{\sum_{i=1}^{n-2}}q_i)
   \Pi^{\text{I},(n)}(p).
   \label{form}
\end{align}
Doing the power counting for the $p$-integration by taking the IR structure of the instanton (\ref{instir}) into account, we see that the first ${\cal{A}}$ field contributes with $\tilde{p}\;p^{-2}$, whereas the last one with $p^{-2}$ for $n \ge 3$ (we have to pick out the most dangerous IR terms) and with $\tilde{p}\;p^{-2}$ for $n=2$. Therefore, we have
\begin{align}
   n &= 2: \qquad 
   p^4 \; (\tilde{p}\;p^{-2}) \; (\tilde{p}\;p^{-2}) \; \tilde{p}^{(2-4)}
   \quad\propto\quad \tilde{p}^0,
   \nonumber
\\
   n &\ge 3: \qquad
   p^4 \; (\tilde{p}\;p^{-2}) \; p^{-2} \; \tilde{p}^{(n-4)} \quad\propto\quad \tilde{p}^{(n-3)}.
\end{align}
The last thing we have to do is to check the remaining integrations over the $q_i$'s. But they are harmless in the IR. With the same argumentation, it can be shown that they are linear in $\tilde{q}_i$. This shows IR finiteness for four-point functions and higher terms. Furthermore, two- and three-point functions can at most lead to logarithmic IR divergences in the instanton background.

Let us start with the contribution of the two-point function (\ref{2}). For the background gauge field ${\cal{A}}$, we insert the non-com\-mu\-ta\-tive instanton (\ref{instir}) given in the previous section. As stated before we are only interested in the IR regime, therefore taking only the small momentum approximation of the instanton solution. 

Using the following results for necessary tensor contractions (remember that we are considering the space $\mathbb{R}_{\text{NC}}^2\times\mathbb{R}^2)$:
\begin{gather}
   \tilde{p}_\mu = \theta_{\mu\nu}p_\nu \quad
   \longmapsto \quad \tilde{p}_1 = \theta p_2, \quad
   \tilde{p}_2 = -\theta p_1, \quad 
   \tilde{p}_3 = \tilde{p}_4 = 0,
   \nonumber
\\
   p^2 = p_1^2+p_2^2+p_3^2+p_4^2, \qquad
   \tilde{p}^2 = \theta^2 (p_1^2+p_2^2),
   \nonumber
\\
   \tilde{p}_\mu\epsilon_{\mu\nu}p_\nu = \theta (p_1^2+p_2^2), \qquad
   \tilde{p}_\mu\epsilon_{\mu\nu}q_\nu = \theta (p_1q_1+p_2q_2),
   \nonumber
\\
   p_\mu\epsilon_{\mu\nu}p_\nu = 0, \qquad
   \epsilon_{\rho\mu}\epsilon_{\rho\nu} = \delta_{\mu\nu},
   \label{tensor}
\end{gather}
we end up with the fairly simple expression
\begin{align}
   S^{\text{I},(2)}_{\text{eff}} = \frac{1}{\pi^2} \int \frac{d^4p}{p^4}.
   \label{log}
\end{align}
Being interested only in the infrared part of this integral, we evaluate it with an ultraviolet cutoff at the instanton size $\Lambda_{\text{UV}}\sim 1/\sqrt{\theta}$ and an infrared cutoff $\Lambda_{\text{IR}}\sim 1/L$ corresponding to a box of size $L$. The result is logarithmically divergent in the IR:
\begin{align}
   S^{\text{I},(2)}_{\text{eff}} = \log \frac{L^2}{\theta}.
\end{align} 
We will postpone a further discussion of this point to the concluding section of this paper. Instead, we go straight to the appropriate calculations of the three-point function.

Applying again (\ref{tensor}) and inserting (\ref{instir}) into (\ref{3}), we get
\begin{align}
   S^{\text{I},(3)}_{\text{eff}} = \frac{\theta^2}{2\pi^4} \int d^4p \; d^4q \;
   \frac{1}{p^2q^2(p+q)^2}
   \left[ p_1q_1+p_2q_2 + \frac{(p_1q_1+p_2q_2)^2}{p_1^2+p_2^2} \right] .
   \label{int}
\end{align}
The structure of this integral suggests a splitting of the four-dimensional momenta into two two-dimensional subparts. In fact, it is clear, considering the underlying integration space $\mathbb{R}_{\text{NC}}^2\times\mathbb{R}^2$, that this procedure makes sense. Let us apply the substitutions
\begin{align}
p=(s,t) \qquad \text{and} \qquad q=(u,v) \,,
\end{align}
to the integral (\ref{int})
\begin{align}
   S^{\text{I},(3)}_{\text{eff}} = \frac{\theta^2}{2\pi^4} \int d^2s \; d^2t \; d^2u \; d^2v \;
   \frac{(s \cdot u)+(s \cdot u)^2/s^2}{(s^2+t^2)(u^2+v^2)((s+u)^2+(t+v)^2)}.
   \label{int1}
\end{align}
Considering the two parts in the numerator separately, calling them $I_1$ and $I_2$, we introduce three Schwinger parameters $\alpha_i$ for the first part; for the second term, we use four of them via the relation
\begin{align}
   \frac{1}{k^2} = \int_0^\infty d\alpha \; e^{-\alpha k^2}.
\end{align}
Representing now the term $(s \cdot u)$ in both integrals with the help of a differential operator, we can perform all 2-dimensional Gaussian integrals over the momenta and end up with
\begin{align}
   I_1 &= - \frac{\theta^2}{2} \int_0^{\infty} d\alpha \; d\beta \; d\gamma \;
   \frac{\gamma}{\tau^3},
\\
   I_2 &= \frac{\theta^2}{4} \int_0^{\infty} d\lambda \; d\alpha \; d\beta \; d\gamma \;
   \frac{\tau+\lambda(\beta+\gamma)+4\gamma^2}
   {\tau(\tau+\lambda(\beta+\gamma))^3},
\end{align}
where we introduced
\begin{align}
   \tau=\alpha\beta+\alpha\gamma+\beta\gamma.
\end{align}
Using Schwinger cutoffs in the UV and IR by implementing factors such as
\begin{align}
   e^{-\frac{1}{\alpha\Lambda_{\text{UV}}^2}-\alpha\Lambda_{\text{IR}}^2},
\end{align}
for every Schwinger parameter, we can perform the integrals $I_1$ and $I_2$ explicitly. The results include only positive powers of $\Lambda_{\text{IR}}$ and are therefore finite in the limit $\Lambda_{\text{IR}}\rightarrow 0$. There is no IR singularity coming from the three-point function, as naive power counting would suggest.
\subsection{The IR Logarithm Structure}
The UV/IR mixing of NC gauge theories makes the coefficient of logarithmic IR divergences that arise from non-planar graphs exactly opposite to that of the logarithmic UV divergences in the planar sector of the theory \cite{Hayakawa:1999yt,Hayakawa:1999zf,Armoni:2000xr,Martin:2000bk}. Therefore, we can write down the corresponding contribution to the effective action in the region of small momenta $p\ll 1/\sqrt{\theta}$\,:
\begin{align}
   S^{\text{II}}_{\text{eff}} ({\cal{A}}) = \frac{1}{2} \int \frac{d^4p}{(2\pi)^4} \;
   \frac{\beta_0}{(4\pi)^2} \log (\Lambda_{\text{UV}}^2\,\tilde{p}^2) \;
   F_{\mu \nu}(-p) \, F_{\mu \nu}(p) + \ldots ,
   \label{ii}
\end{align}
where $\beta_0=11/3$ is the one-loop beta-function coefficient for NC $U(1)$ YM and the ultraviolet cutoff is again given by $\Lambda_{\text{UV}}\sim 1/\sqrt{\theta}$. The dots denote higher terms, which have to be implemented in order to render the effective action gauge-invariant. We must use again a generalization of open Wilson lines \cite{Armoni:2001uw}, but for our purpose the leading part is sufficient.

Let us first calculate the contribution of the two-point function to the effective action and then give arguments why higher-point functions are irrelevant to IR divergences. We rewrite (\ref{ii}) as
\begin{align}
   S^{\text{II}}_{\text{eff}} ({\cal{A}}) =
   \int \frac{d^4p}{(2\pi)^4} \; {\cal{A}}_{\mu}(p) {\cal{A}}_{\nu}(-p) \Pi_{\mu\nu}^{\text{II},(2)}(p)
   +{\cal{O}}({\cal{A}}^3),
   \label{ii2}
\end{align}
with
\begin{align}
   \Pi_{\mu\nu}^{\text{II},(2)}(p) &= -
   \frac{\beta_0}{(4\pi)^2} \log (\Lambda_{\text{UV}}^2\,\tilde{p}^2)
   \; (p^2\delta_{\mu\nu}-p_{\mu}p_{\nu}).
   \label{ii22}
\end{align}
Inserting in (\ref{ii2}) the instanton field (\ref{instir}) and applying the relations (\ref{tensor}) yields the following integral for the leading part of (\ref{ii2}):
\begin{align}
   S^{\text{II},(2)}_{\text{eff}} = \frac{\beta_0}{(4\pi)^2} \; \theta^2 \int d^4p \;
   \log ((p_1^2+p_2^2)\theta).
\end{align}
Splitting again the four-di\-men\-sional momentum space into two two-di\-men\-sional parts via $p=(s,t)$ leads to
\begin{align}
   S^{\text{II},(2)}_{\text{eff}} = \frac{\beta_0}{(4\pi)^2} \; \theta^2 \int d^2s \; d^2t\;
   \log (s^2 \theta) .
\end{align}
Making use of the formula
\begin{align}
   \log\alpha = - \int_1^\infty \frac{e^{-\alpha\beta}}{\beta} d\beta - \gamma_E + {\cal{O}}(\alpha),
\end{align}
and performing the Gaussian momentum integral shows that this expression is completely finite in the IR. There are no IR divergences arising from the logarithmic piece (\ref{ii22}) of the two-point function.

What happens in the case of higher-point functions? They are all safe in the IR regime ($\tilde{p}_i\rightarrow 0$) since, because of the $\star$-product, every $\log$-term gets multiplied by $\sin(\tilde{p}_ip_j/2)$, which renders the whole expression finite.
\subsection{Supersymmetry as an IR Regulator}
It is a well-known fact that supersymmetric non-com\-mu\-ta\-tive field theories do not show any quadratic or linear UV/IR mixing \cite{Matusis:2000jf,Zanon:2000nq,Ruiz:2000hu}. Therefore, we expect that the instanton calculation outlined here will be better behaved in the supersymmetric case.

To demonstrate this, we consider a theory with one gauge field, $n_f$ Weyl fermions and $n_s$ real scalars, where we take again the simplest gauge group $U(1)$. The leading terms of the non-planar two-point function of the gauge field in the IR regime are given by (which would replace (\ref{pii2}) above)
\begin{align}
   \Pi_{\mu\nu}^{\text{I},\text{SUSY}}(p) &= -
   \frac{1}{\pi^2} \left(1-n_f+\frac{n_s}{2}\right) \frac{\tilde{p}_{\mu}\tilde{p}_{\nu}}{\tilde{p}^4}
   +{\cal{O}}(\tilde{p}^0).
   \label{piisusy}
\end{align}
The expression in brackets is always zero for a theory with supersymmetric field content. Because for ${\cal{N}}=1$ NCSYM we have $n_f=1$, $n_s=0$, whereas in the case of ${\cal{N}}=2$ we have $n_f=n_s=2$, and finally for ${\cal{N}}=4$ we have to apply $n_f=4$ and $n_s=6$. This cancellation between bosonic and fermionic modes also takes place for higher-point functions. Therefore, no IR divergences come from the pole-like structure of non-planar supersymmetric $n$-point functions.

Next, we have to generalize expression (\ref{ii}) to the supersymmetric case. This is again achieved very easily in replacing the coefficient of the beta-function by
\begin{align}
   \beta_0 = \frac{1}{3} \left(11-2n_f-\frac{n_s}{2}\right).
\end{align} 
In the case of non-com\-mu\-ta\-tive ${\cal{N}}=4$ theory, there is still a vanishing beta-function, whereas we get logarithmic non-planar corrections for ${\cal{N}}=2$ and ${\cal{N}}=1$ NCSYM. Nevertheless, as shown above, they are harmless with respect to the instanton determinant.

A last comment should be made for softly broken supersymmetric theories. These theories have different masses for the fermions and scalars, which partly breaks supersymmetry. But these soft breaking effects do not change the leading IR structure of non-planar $n$-point functions.\footnote{Remarkably, subleading terms can lead to tachyonic modes of the photon in Lorentzian signature \cite{Landsteiner:2000bw,Landsteiner:2001ky,Carlson:2002zb,Alvarez-Gaume:2003mb}.} The instanton determinant is again IR-safe in these theories.

Therefore, we conclude that supersymmetry helps in rendering the one-loop instanton determinant IR-finite, whereas in the non-supersymmetric case we find a logarithmic divergence.

\section{Conclusions}\label{con}
We have studied the impact of UV/IR mixing on the one-loop instanton determinant of non-com\-mu\-ta\-tive $U(1)$ Yang--Mills theory. The anti-selfdual instanton has a classical size of order $\sqrt{\theta}$ and decreases with $1/r^3$ away from the origin. It is therefore well behaved in the classical approximation of the theory.

Taking one-loop quantum effects into account, we find a logarithmic infrared divergence of the instanton determinant coming from the non-planar two-point function. {\em Non-com\-mu\-ta\-tive quantum fluctuations blow the classical finite size of the instanton up to infinity.}

It should be clarified here that the {\em blow up\/} of the instanton size to infinity is a metaphor, a way of conveying the message that at the end, the non-com\-mu\-ta\-tive Yang--Mills theory behaves similarly to the ordinary non-Abelian Yang--Mills theory, having an IR-divergent dilute instanton measure, although for different reasons when it comes to the details. In the case of ordinary Yang--Mills theory, the IR problem comes from the integral over instanton sizes, whereas in the non-com\-mu\-ta\-tive case it is because of the UV/IR mixing effects.

Hence, dilute instanton calculus in non-com\-mu\-ta\-tive Yang--Mills theory is ruined by the infrared catastrophe, unless we work in finite volume (see \cite{Alvarez-Gaume:2001tv} for a discussion of such a case in a related context) or we work in softly broken supersymmetric theories.
\section*{Acknowledgements}
The author would like to express his gratitude to Jos{\'e} L.~F.~Barb{\'o}n for numerous enlightening discussions and careful reading of the manuscript.
\end{document}